\documentclass[final,3p,times,twocolumn]{elsarticle}
\usepackage{graphicx}
\usepackage{epsfig}
\usepackage{amssymb}
\usepackage{amsmath}
\usepackage{color}
\usepackage{float}
\journal{Physica C}

\begin{document}

\begin{frontmatter}

\title{Anisotropic behaviour of transmission through thin superconducting NbN film in parallel magnetic field}

\author[FZU]{M. \v{S}indler} \ead{sindler@fzu.cz}
\author[FZU,MFF]{R. Tesa\v{r}}
\author[FZU]    {J. Kol\'{a}\v{c}ek}
\author[MFF] {L.~Skrbek}

\address[FZU]{Institute of Physics ASCR, v. v. i., Cukrovarnick\'{a} 10, CZ-162 53 Praha 6, Czech Republic}
\address[MFF]{Faculty of Mathematics and Physics, Charles University, Ke Karlovu 3, CZ-121 16 Praha, Czech Republic}

\date{\today}

\begin{abstract}
Transmission of terahertz waves through a thin layer of the superconductor NbN  deposited on an anisotropic R-cut sapphire substrate is
studied as a function of temperature  in a magnetic field oriented parallel with the sample. A significant difference is found between transmitted intensities of  beams  linearly  polarized parallel with and perpendicular to the direction of applied magnetic field.

\end{abstract}


\end{frontmatter}

\section{Introduction}

\begin{figure}
\includegraphics[width=0.45\textwidth]{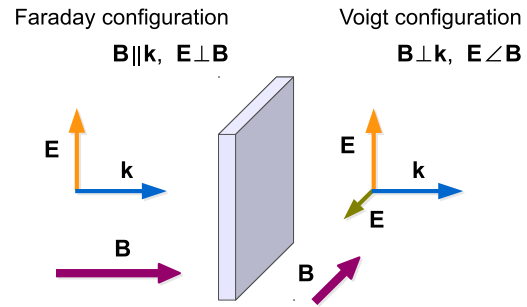}
\caption{ Schematics of fundamental magneto-optical geometries: $\vec{B}$ is a static external magnetic field, $\vec{k}$ is a wavevector and $\vec{E}$ denotes electric field of an incoming beam.}
\label{mgo_configs}
\end{figure}

Far-infrared techniques provide  valuable information about condensate, quasiparticle and vortex behavior in both equilibrium and non-equilibrium states of superconductors \cite{dressel_overview}. Magnetic fields substantially change the properties of superconductors in two different ways: by pair-breaking mechanism or by introducing a vortex lattice.
Regarding mutual orientation of applied magnetic field, sample position and linear polarisation, three fundamental configurations of magnetooptical experiments can be recognised, as illustrated in figure~\ref{mgo_configs}.
While in the \emph{Faraday geometry}, optical properties are insensitive to the direction of the incident beam polarisation, in the \emph{Voigt geometry}, the dependence is generally possible.

As the case of Faraday geometry is better understood \cite{ikebe09,Xi_vortex,sindler2014} and only few studies deal with the parallel field \cite{luzhbin_structure,xi,Grbic}, in this work we concentrate on transmission in Voigt geometry.
Very thin superconducting films of thickness, $d$,
 lower or comparable with the coherence length, $\xi$, exhibit in parallel magnetic field a  vortex free state~\cite{xi}. In thicker films, vortices are introduced inside the film. Luzhbin \cite{luzhbin_structure} found that in such a geometrically frustrated system vortices arrange themselves in rows.

In this paper, we describe transmission of monochromatic THz radiation with well defined linear polarisation through a thin NbN film with magnetic field applied parallel to its surface. Anisotropy of transmission can be regarded as a sign of vortex presence.

\section{Experiment}
The superconducting NbN film of nominal thickness $d_1 = 15$~nm
was grown epitaxially on an anisotropic R-cut sapphire substrate (with the c-axis at angle $\varphi=57.6^{\circ}$ with respect to the surface normal), see table \ref{vzorek}. This sample has already been studied by us
in zero magnetic field \cite{tesar2011}.

\begin{table}[h]
\centering \caption{Parameters of the sample} \label{vzorek}
\begin{tabular}{|c|c|c|c|c|c|c|}
\hline
Film &  $d_{1}$ &  $T_c$   & $\sigma_N(0)$ &substrate & $d_{2}$ \\
\ &  [nm]  & [K] &  $\Omega ^{-1}.m^{-1}$ &\ & [mm]  \\
\hline
NbN  & 15 & 16  & $0.51 \times 10^{6}$ & Al$_2$O$_3$ & 0.33  \\
\hline
\end{tabular}
\end{table}

Our optically pumped far-infrared laser produces several discrete lines in the range
from 0.4 to 4.9~THz. Intensity of transmitted beam through the sample placed
inside magneto-optical cryostat is measured by a helium cooled bolometer
and relative transmission is evaluated as a ratio between measured transmission and transmission measured just above $T_c$. Our experimental protocol is to set a constant magnetic field  and to cool the sample from slightly above $T_c$ to the minimum attainable temperature (usually $3$ K); then the sample is heated up. Temperature of the sample is monitored by a small Cernox thermometer located nearby, which performs well even in high magnetic fields. The DC resistivity is monitored  simultaneously, by the four-probe measurement method. Further details can be found elsewhere~\cite{tesar2010}.

Transmission is measured for horizontal and vertical linear polarisations utilizing the wire-grid polarisation. In our special case, the angle 
between the electric vector of the incident beam and the extraordinary ray axis of the birefringent sapphire substrate is  $45^{\circ}$, therefore 
the relative transmissions of the horizontally and vertically polarized radiations should be equal in zero magnetic field.
Note, however, that the intensities of the horizontally and vertically polarised rays are different, which might affect the accuracy of the individual measurements, but resulting dependencies are reliable.

\section{Theoretical model}
Zero magnetic field conductivity $\tilde{\sigma}_s$ is well described by the BCS theory-based Zimmermann model~\cite{Zimmermann}. In our model, vortices are considered as cylindrical inclusions of normal
state material~\cite{clem75}. Their typical radius is proportional to the coherence length ($\xi \cong 5$~nm for
NbN at zero temperature) which is much smaller than the typical wavelengths in the
terahertz range. We can therefore use the long wavelength limit when electromagnetic
radiation does not sense individual vortices and the system can be thought of as
a homogeneous one, possessing an effective complex conductivity~$\tilde{\sigma}_{eff}$ which strongly depends on the
volume fraction of vortex cores $f_n = V_n/V$ and the superconducting fraction $f_s=1-f_n$.

In our case, vortices are considered as inclusions and the superconducting environment surrounding the vortices as a filler.
Maxwell-Garnett formulated his theory~\cite{MGT} for dilute systems assuming that inclusions feel the external field that coincide with the local field inside 
the filler.
In contrast with the Bruggeman theory~\cite{Bruggeman}, this model respects the topology of the vortex system.
Mutual interaction between inclusions is neglected. More generally, it was found that MGT formulas can hold even for higher concentrations of inclusions 
as long as the filler is percolated \cite{rychet}.
Considering the special case of cylindrical inclusions,
for electric field perpendicular and parallel to the vortex axis, MGT gives 
\begin{equation}
 \tilde{\sigma}_{MGT}^{\perp}=\frac{2 f_n \tilde{\sigma}_s (\tilde{\sigma}_n - \tilde{\sigma}_s)}{(1-f_n)(\tilde{\sigma}_n - \tilde{\sigma}_s)+2\tilde{\sigma}_s}+\tilde{\sigma}_s\,\,;
\end{equation}

\begin{equation}
 \tilde{\sigma}_{MGT}^{\parallel}=f_n \tilde{\sigma}_n +(1-f_n) \tilde{\sigma}_s\,\,,
\label{eps_paralel}
\end{equation}
for electric fields perpendicular to and parallel with the vortex axes, respectively.
The latter formula can be qualitatively understood assuming that charges along the vortex cores and in the superconducting matrix move independently.

\section{Results}

Zero-field transmission has been measured in both polarisations. The observed difference does not exceed 5\%, which most likely can be attributed to a slight misalignment of the sample.
Importantly, one can appreciate an agreement between transmission obtained while cooling and heating our sample.
The calculations for both vertical and horizontal polarisations are performed using the following values of physical properties of our sample: $\tau=3.86$ fs \cite{semenov} and $2\Delta(0)=5.7$~meV which results in $2\Delta(0)/ k_B T_c = 4.15$  ratio~\cite{tesar2011}.

While variations of the film parameters within their experimental errors affect the calculated transmission only slightly, the results are much more sensitive to the exact values of ordinary and extraordinary refractive indices of the substrate. This is well understood, since transmission is influenced by the interference effects.

\begin{figure}[H]
\centering
\includegraphics[height=9cm]{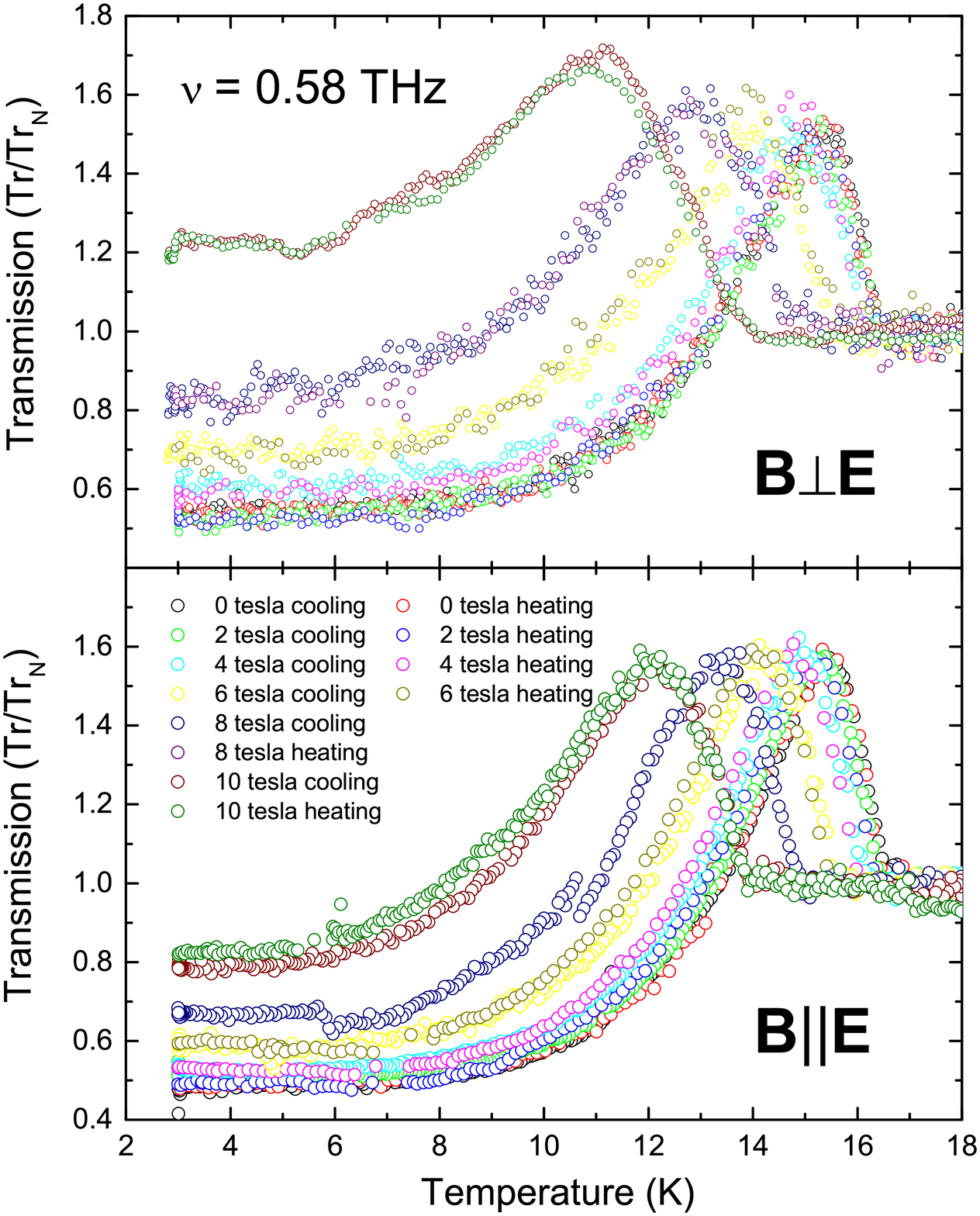} 
\caption{Temperature dependence of transmission through NbN thin film at 0.58~THz in parallel magnetic fields up to 10~T. Upper panel: perpendicular polarisation; lower panel: horizontal polarisation.}
\label{voigt513um}
\end{figure}

\begin{figure}[h]
\centering
\includegraphics[height=9cm]{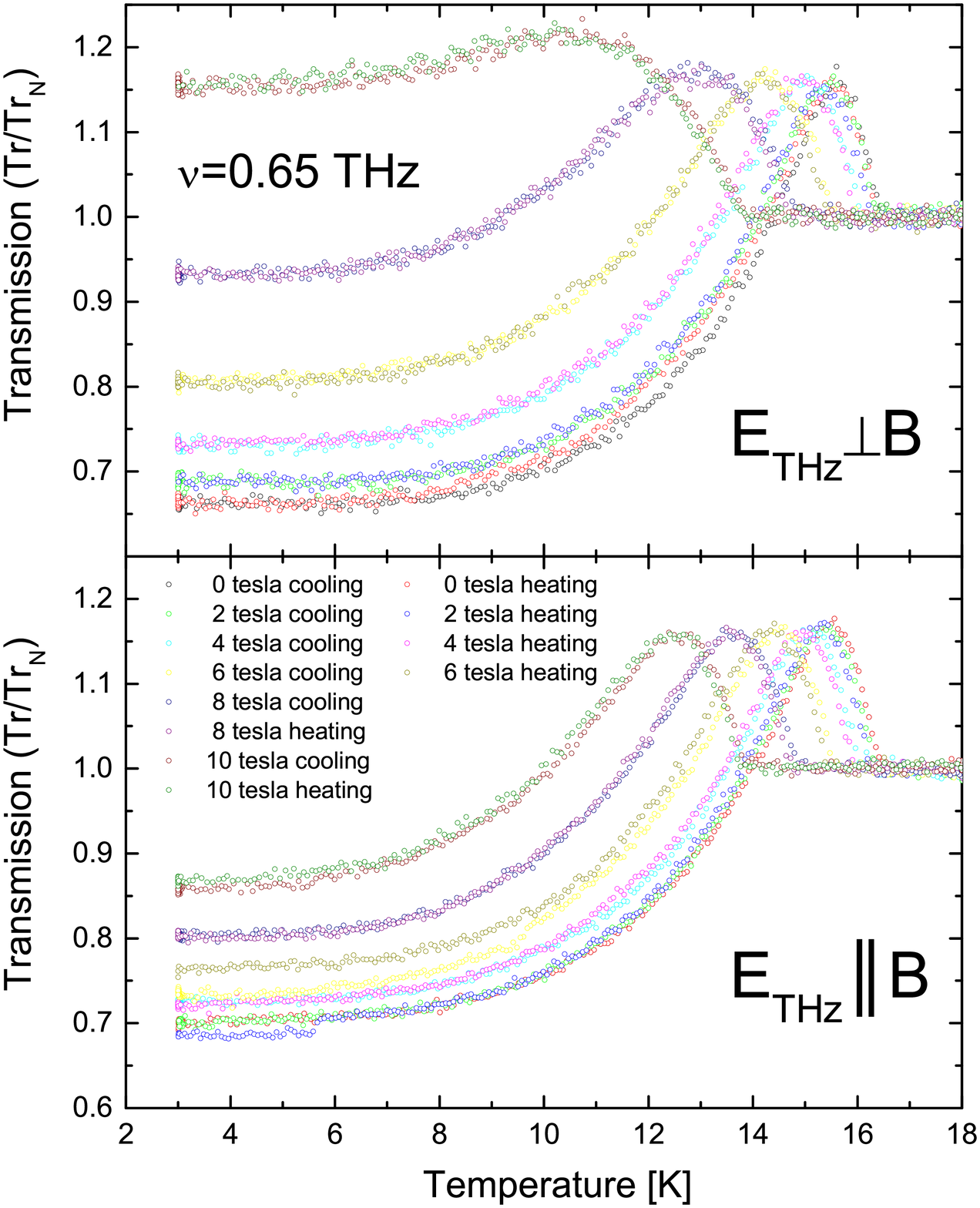}
\caption{Temperature dependence of transmission through NbN thin film at 0.65~THz in parallel magnetic fields up to 10~T. Upper panel: perpendicular polarisation; lower panel: horizontal polarisation.}
\label{voigt458um}
\end{figure}

Let us describe our experimental results. Transmission is measured at frequencies 2.52, 0.65, 0.58, 0.53 and 0.40~THz. In low magnetic fields up to 2 T, the measured transmission deviates from the zero field values only slightly, usually within the experimental error.
In higher fields, the dependence of transmission on magnetic field and on the direction of linear polarisation becomes clearly visible.
Only for the lowest frequency (0.40~THz), we find transmission insensitive to the polarisation direction.
In large magnetic fields, significant differences in transmission of rays with different polarisation is observed in the case of 0.58~THz (figure \ref{voigt513um}) and  0.65~THz (figure \ref{voigt458um}) lines. For the 0.53~THz and 2.52~THz lines anisotropy is less pronounced,  these experimental results are therefore not shown.

\begin{figure}[ht]
\includegraphics[height=9cm]{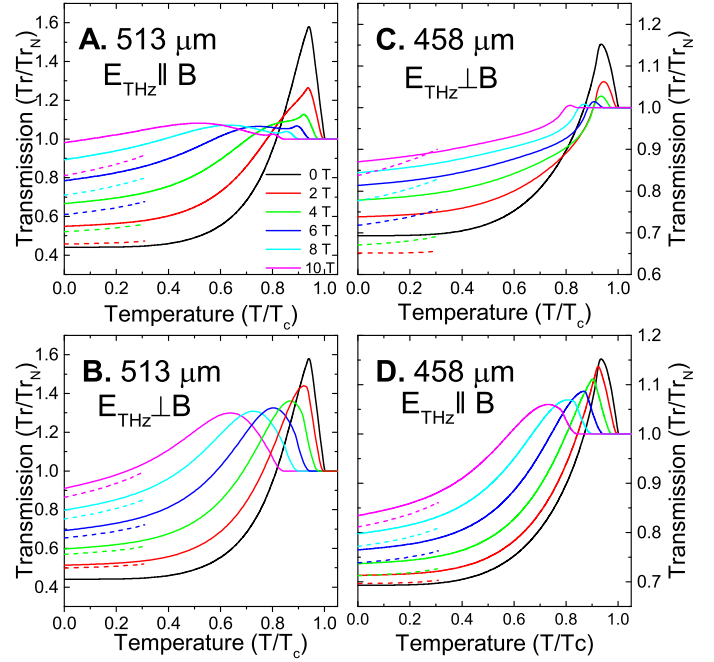}
\caption{Numerical simulation of transmission through a thin NbN film in Voigt geometry. For details, see the text.}
\label{voigt_theory}
\end{figure}

\section{Discussion}
In non-zero magnetic field, both NbN film and R-cut sapphire are anisotropic, which leads to new features. It is no longer possible to decompose the beam into two different parts with their own eigenpolarisations - mode coupling occurs \cite{yeh79,yeh80,sindler2012}. This prevents us to treat transmission of linearly polarised beam parallel to and perpendicular with applied magnetic field independently.

The key variable in our theoretical model is the vortex core fraction $f_n$
which can be approximated by
$f_n = B/B_{c2}(T)$~\cite{ikebe09}.
As shown by Drew \cite{drew09}, the time dependent Ginzburg Landau theory predicts the upper critical field in the thin film limit in parallel field  as $B_{c2}=B_{c2}(0) \sqrt{1-t}$. Our results are well-described by this Drew formula with $B_{c2}(0)=25.1$~T
and almost equally well by the equation $B_{c2}=B_{c2}(0) \sqrt{1-t^2}$ with $B_{c2}(0)=18.4$~T.
The normal state conductivity of vortex cores,  $\tilde{\sigma}_n$, is considered to be independent of magnetic field.

Numerical calculations of the transmission displayed in fig.~\ref{voigt_theory} are performed following our method as described in Ref.~\cite{sindler2012} using values given in table~\ref{vzorek} and the Drew formula for $f_n$. In our previous work~\cite{sindler2012}, we used the Bruggeman approach to predict the behaviour of transmission in the Voigt geometry, however, the MGT approach used here describes the underlying physics of the problem better.
Indeed, the resulting theoretical curves describe our transmission data for the beam polarized parallel with vortex axis semi-quantitatively. 
With increasing magnetic field, the thermal peaks diminish and the almost constant transmission at low temperatures shifts towards higher values. The experimentally observed peaks are somewhat broader then predicted by the theoretical calculations, which can be attributed to some inhomogeneity of our sample.

In the case of perpendicular polarisation, the effective medium approximation fails to describe the experimentally observed transmission, even on a qualitative level. The reason might be that, as mentioned above, the model does not account for vortex motion.
At low temperatures, vortex dynamics is described by the Coffey-Clem model~\cite{sindler2014} applicable for Faraday geometry. We applied the same approach 
for our case, see dashed lines in fig.~\ref{voigt_theory}. Although the effect of Coffey-Clem contribution is clearly appreciable, it does not account for the
observed anisotropy satisfactorily.
We note that, due to the mode coupling, vortex dynamics influence transmission even in the case of electric field parallel with the vortex axis.
\section{Conclusions}
In the Voigt experimental geometry, transmission of beams with  polarisation parallel with and
perpendicular to the vortex axes in the superconducting thin NbN film exhibit very different behaviour.
In the case of parallel polarised beams, we report a fairly good
semi-quantitative description of transmission, while for perpendicularly polarised beams, the Maxwell-Garnett theory
fails to describe measured transmission on the qualitative level.
We believe that the simple formula~\ref{eps_paralel} provides good approximation of conductivity in the direction parallel with the applied magnetic field $\tilde{\sigma}^{\parallel}$, while conductivity in perpendicular direction
$\tilde{\sigma}^{\perp}$ is more complex and deserves further attention.
We therefore plan systematic measurements of transmission through NbN
films deposited on MgO isotropic substrate utilizing laser thermal spectroscopy
and time-domain terahertz spectroscopy methods for both types of Voigt geometry,
which will be reported elsewhere.

\section*{Acknowledgments}
We are grateful to K.~Il'in and M.~Siegel for preparing the NbN film,
 P. Ku\v{z}el and C. Kadlec for
measurement of the refractive indices of the substrate and to
J. \v{S}ebek, J. Prokle\v{s}ka and M. \v{Z}\'{a}\v{c}ek for the dc conductivity
measurements of our NbN film. We acknowledge support by the MSMT grant \#LD14060 and  the
COST action MP1201.


\end{document}